%% file: Poincare-index-BEC-final.tex
\documentclass[a4paper,11pt]{article}
\pdfoutput=1 

\usepackage{jheppub} 

\usepackage[english]{babel}	
\makeatletter\AtBeginDocument{\let\@elt\relax}\makeatother 
\usepackage{amsmath}		
\usepackage{amssymb}		
\usepackage{bbm}			
\usepackage{array}			
\usepackage{graphicx}		
\usepackage{graphics}		
\DeclareGraphicsExtensions{.jpg,.jpeg,.pdf,.png,.eps}	

\newcommand{\bs}{\boldsymbol}
\newcommand{\Figref}[1]{Fig.~\ref{#1}}
\newcommand{\Eqref}[1]{\eqref{#1}}

\newcommand{\Grad}{{\bs \nabla}}

\newcommand{\x}{\mathbf{x}}

\newcommand{\ez}{{\bs e}_z}
\newcommand{\vv}{\mathbf{v}}
\newcommand{\vw}{\mathbf{w}}
\newcommand{\vj}{\mathbf{j}}

\newcommand{\la}[1]{\label{#1}}

\title{\boldmath Poincar\'e index formula and analogy with the 
Kosterlitz-Thouless transition in a non-rotated  cold atom Bose-Einstein 
condensate}

\author[a,b,1]{Julien Garaud\,\note{Corresponding author.}} 
\author[b,c]{Antti J. Niemi\,} 

\affiliation[a]{Institut Denis Poisson CNRS-UMR 7013, Universit\'e de Tours, 37200 France}
\affiliation[b]{Nordita, Stockholm University and Uppsala University, \\  
				Roslagstullsbacken 23, SE-106 91 Stockholm, Sweden}
\affiliation[c]{Department of Physics, Beijing Institute of Technology, 
				Haidian District, Beijing 100081, China}
\emailAdd{garaud.phys@gmail.com}
\emailAdd{Antti.Niemi@su.se}

\abstract{
A dilute gas of Bose-Einstein condensed atoms in a non-rotated and axially symmetric 
harmonic trap is modelled by the time dependent Gross-Pitaevskii equation.  
When the angular momentum carried by the condensate does not vanish, the minimum 
energy state describes vortices (or antivortices) that propagate around the trap center. 
The number of (anti)vortices increases with the angular momentum, and they repel each other 
to form Abrikosov lattices. 
Besides vortices and antivortices there are also stagnation points where the superflow vanishes; 
to our knowledge the stagnation points have not been analyzed previously, in the context of 
the Gross-Pitaevskii equation. The Poincar\'e index formula states that the difference 
in the number of vortices and stagnation points can never change.
When the number of stagnation points is small, they tend to aggregate into degenerate 
propagating structures.  
But when the number becomes sufficiently large, the stagnation points tend to pair up 
with the vortex cores, to propagate around the trap center in regular lattice arrangements.  
There is an analogy with the geometry of the Kosterlitz-Thouless transition, with
the angular momentum of the condensate as the external control parameter instead 
of the temperature.
}

\arxivnumber{2108.03155}

\begin{document} 
\maketitle
\flushbottom
\section{Introduction}
\label{Sec:Intro}

A Bose-Einstein condensate is a coherent macroscopic quantum state with unique features 
that facilitate a high level of experimental control \cite{Anderson-1995,
Davis-1995,Bradley-1997}. In particular condensates that form in a trapped, rotated gas of 
ultra-cold alkali atoms are studied extensively, both in earth-bound and in earth-orbiting 
laboratory experiments \cite{Aveline-2020}. Among the emerging applications is the 
development of ultra-sensitive sensors and detectors \cite{Elliott-2018}. The properties 
of cold atom condensates are also under active investigation, as a potential platform 
for quantum computation and simulation \cite{Bloch-2012}.
At the level of the mean field theory a condensate can be modelled by a macroscopic wave 
function that solves the time-dependent Gross-Pitaevskii equation \cite{Gross-1961,
Pitaevskii-1961}. This is a nonlinear Schr\"odinger equation with a quartic nonlinearity 
that accounts for the interactions between atoms \cite{Pitaevskii-2003,Lieb-2006,
Pethick-2008,Fetter-2009,Bao-2013}. Quantum vortices are the principal topological 
excitations of a trapped cold atom Bose-Einstein condensate \cite{Matthews-1999,Chevy-2000,
Raman-2001,Abo-Shaeer-2001,Sonin-2016}. They have been studied extensively in terms of 
solutions of a stationary Gross-Pitaevskii equation that models a uniformly rotating 
condensate \cite{Butts-1999,Aftalion-2001,Haljan-2001,Fetter-2001,Fetter-2009}, for detailed 
reviews, see e.g., \cite{Aftalion-2006,Malomed-2019}).  

The time-dependent Gross-Pitaevskii equation appears in a broad range of physical systems, 
beyond condensed matter physics. For example, in combination with a gravitational trapping 
potential it can model Dark Matter as a Bose-Einstein condensate \cite{Boehmer.Harko-2007,
Nikolaieva.Olashyn.ea-2021}. 
Using the Madelung transformation, the equation can be sent into a system of equations that 
describe a barotropic-type fluid, which can model codimension two membranes in higher 
dimensions~\citep{Khesin.Yang-2021}.   
Furthermore, in  two and three space dimensions and in the limit of constant and uniform density,  
one arrives at the fluid dynamical Euler equation that supports pointlike vortex filaments, 
with dynamics that is described by the localized induction approximation \cite{Newton}.
Finally, in two dimensions these point vortices also appear in an effective field 
theory that evaluates a conformal field theory operator spectrum at large global 
charge~\cite{Cuomo.Fuente.ea-2018}.

Topological methods can be highly effective in the characterization of a physical system, 
when the goal is to identify and describe robust phenomena. The numerous applications 
range from fundamental interactions to condensed matter, fluid dynamics and beyond 
\cite{Nakahara}. 
Here we extend the repertoire, in the case of a cold atom Bose-Einstein condensate, 
by the Poincar\'e index formula \cite{Hartman-1964,Llibre-1998,Lloyd-1978}. It is 
a refinement of the Poincar\'e-Hopf index theorem \cite{Milnor-1965} that is used 
widely to analyze topological aspects of vectors fields.
We employ the Poincar\'e index formula to identify and analyze previously unreported 
properties of quantum vortices in the context of cold atom Bose-Einstein condensates.  
In particular, we identify the presence of a changeover transition that displays a 
geometry that is quite similar to the geometry of vortices and antivortices in the 
Kosterlitz-Thouless transition that occurs {\it e.g.} in the two dimensional XY-model 
and in the plane rotor model  \cite{Kosterlitz-1972,Kosterlitz-1973,Nobel}.
Instead of the temperature, now the external control parameter is the internal angular
momentum that is supported by the condensate; in the case of a non-rotating trap its 
value can be regulated externally, possibly by optical photon beams.

A vortex in the two dimensional Gross-Pitaevskii model has a pointlike core where the 
modulus of the wave function vanishes. The energy of a trapped vortex is finite, albeit 
logarithmically increasing in the length scale that characterizes the size of the trap.  
The vortex is a topological construct with an integer valued {\it circulation} of the 
superflow velocity $\vv(\x,t)$ around its core; in the case of a single vortex core the 
circulation has the normalized value  $\pm 1$ where the sign depends on the orientation
of the circulation, and the total circulation of a multivortex configuration is the sum 
of its individual vortex circulations. Commonly the circulation around the vortex core 
is the only topological invariant that is assigned to a vortex system, in the case of a 
cold atom Bose-Einstein condensate. 

Here we consider the free energy minima of the two dimensional Gross-Pitaevskii model, 
in the case of a nonrotating harmonic trap. We observe that besides the circulation, 
there are also other topological structures that can be assigned to a multivortex 
configuration. In particular, besides the vortex cores that are topologically centers 
of the vector field $\vv(\x,t)$, there can also be saddles of $\vv(\x,t)$. These are 
points where the flow velocities around distinct vortices exactly cancel each other, 
and thus where the superflow vector field stagnates. Unlike in the case 
of a vortex core, the density of the condensate does not vanish at a stagnation point. 
Nevertheless, the stagnation points are also topological structures that can not be 
removed by a generic local deformation of the wave function. Moreover, unlike in the 
case of a vortex core the circulation of $\vv(\x,t)$ can not directly detect the presence 
of a stagnation point. But all the fixed point structures of $\vv(\x,t)$ can be 
assigned an integer valued {\it winding number} that is a topological invariant with a 
content that can be summarized by the Poincar\'e  index formula 
\cite{Hartman-1964,Llibre-1998,Lloyd-1978}.

In the common presentation of the Kosterlitz-Thouless transition in the case of 
the XY-model and the plane rotor model  \cite{Nobel}, an antivortex has the topology 
of a saddle; see Appendix for a detailed description. But to our knowledge the properties 
of the saddles where the superflow stagnates have not been previously analyzed, 
in the case of cold atoms condensates. Accordingly we apply the Poincar\'e index 
formula to analyze both the vortex cores and the superflow stagnation  points. To illustrate 
this, we use a particular set of solutions of the Gross-Pitaevskii equation, those with 
a minimal energy for fixed values of the particle number and angular momentum, but in a 
trap that does not rotate. Our methods and results should help to clarify the different 
role of the circulation and the winding number as complementary integer valued topological 
invariants, to characterize a multivortex structure in the case of a Bose-Einstein 
condensate.
For this we start and introduce the Poincar\'e index formula that is relevant for the 
description of the topology of the velocity superflow in two spatial dimensions. 
We apply this formula to analyze a set of multivortex solutions of the Gross-Pitaevskii 
equation that describes a dilute, non-rotating condensate of cold atoms in an harmonic 
trap. In particular, we observe and describe in detail, using the Poincar\'e index 
formula, how the geometric pattern of vortex cores and stagnation points undergo a transition 
that is analogous to the  Kosterlitz-Thouless transition of vortices and antivortices 
{\it e.g.} in the common presentation of the XY-model, but with the value of angular 
momentum as the external control parameter instead of temperature; in the Appendix 
we compare the topological structures, as they commonly appear in the two cases.

\section{Theoretical framework}
\label{Sec:Theory} 

\subsection{Circulation}
\label{Sec:circulation}

In a theoretical approach, a trapped Bose-Einstein condensate of cold atoms is commonly 
modelled by a macroscopic, complex-valued and normalizable wave function $\psi(\x,t)$. 
The modulus $|\psi(\x,t)|$ describes the density of the condensate and its phase determines 
the velocity vector field of the superflow
\begin{equation}
\vv(\x,t) = \Grad\arg[\psi] (\x,t)	\,.
\label{v}
\end{equation}
The principal topological excitations in these condensates are vortices \cite{Sonin-2016}.
In a two dimensional model, the core of a vortex is a center of the superflow velocity 
$\vv(\x,t)$; at the vortex core the modulus of the wave function vanishes. As a topological 
invariant, a vortex is commonly characterized by the integer valued circulation 
\begin{equation}
n_\vv(p;\Gamma) = \frac{1}{2\pi} \oint_\Gamma d{\bs\ell} \cdot \vv 
 \in \mathbb Z \,,
 \la{winding}
\end{equation}
of the superflow velocity around its core. For a simple closed, counterclockwise curve 
$\Gamma$ a single vortex core contributes $\pm 1$ to the circulation \Eqref{winding}, with 
sign depending on the direction of $\vv(\x,t)$ along the curve; the right-hand panel 
of Figure \ref{Fig:Fixed-points} shows an example of a vector field $\vv(\x)$ with 
$n_\vv(p;\Gamma) = +1$. In the case of several vortex cores in the interior of 
the curve, the circulation \Eqref{winding} counts their net number.   

In addition of vortex cores which are centers of the vector field 
$\vv(\x,t)$, there are also other topologically invariant fixed points that can be 
assigned to a two dimensional vector field including sinks, sources and saddles.
From the knowledge of all its fixed points, a phase portrait of the vector field 
$\vv (\x,t)$ can then be constructed. It characterizes the entire vector field 
$\vv (\x,t)$, in a manner that is invariant under continuous local deformations. 
A phase portrait commonly describes adequately the solutions of the underlying dynamical 
equation, often without any need to actually solve the equation.

\subsection{Winding number}
\label{Sec:Wind}

\begin{figure*}[!tb]
\hbox to \linewidth{ \hss
\includegraphics[width=\linewidth]{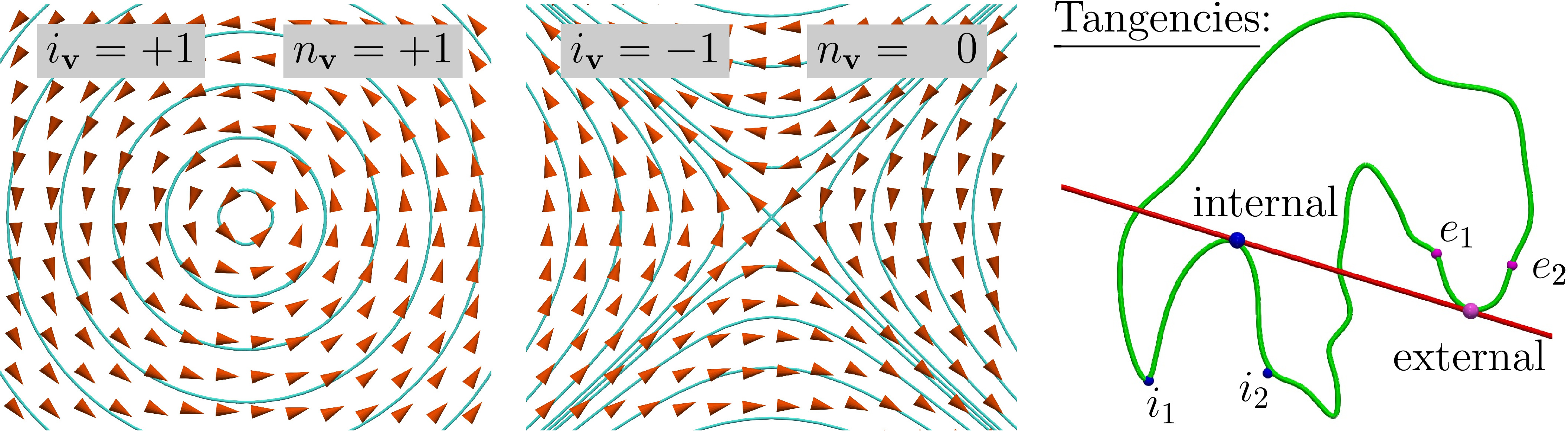}
\hss} 
\caption{ 
Phase portraits that are encountered in the present study of the superflow velocity 
vector $\vv(\x)$  in \Eqref{v}. 
The red arrows show the vector field $\vv(\x)$, and the cyan lines represent its 
streamlines.
The first two panels, respectively, show a center that corresponds to a 
single vortex core with $n_\vv = 1$ and $i_\vv=1$, and a saddle  that corresponds to
a stagnation point with $n_\vv = 0$ and 
$i_\vv=-1$. 
The rightmost panel illustrates what are internal and external tangencies.
Since the curve segment between $i_1$ and $i_2$ is concave, it ads one to $I_\Gamma$.  
Likewise, the curve segment between $e_1$ and $e_2$ is convex and it ads one to 
$E_\Gamma$. It is notable that both centers and saddles appear in the common
description of  Kosterlitz-Thouless theory of phase transitions in the 2D XY model. 
In that model vortex cores appear topologically as centers and antivortex cores appear 
as saddles, and both have an energy that diverges logarithmically when the short 
distance cutoff goes to zero \cite{Nobel} (See Appendix.)
}
\label{Fig:Fixed-points}
\end{figure*}

In addition of vortex cores that are mathematically centers of the vector field $\vv (\x,t)$, 
stagnation points of $\vv (\x,t)$ can also be present in cold atom Bose-Einstein condensates.   
Stagnation points are mathematically saddles of $\vv (\x,t)$. At a stagnation point the 
superflow vanishes as the velocity \Eqref{v} of distinct vortices exactly cancel each other 
while the density of the condensate does not need to vanish. But unlike centers, saddles 
can not be captured by the circulation \Eqref{winding}.  For example, the middle panel of 
Figure \ref{Fig:Fixed-points} shows a vector field with a saddle. The index \Eqref{winding} 
vanishes when evaluated over any simple counterclockwise  curve $\Gamma$ that encircles the 
saddle.

Besides the circulation \Eqref{winding}, as a differentiable two dimensional vector 
field, the superflow velocity \Eqref{v} also supports the integer valued winding 
number
\begin{equation}
i_\vv(p;\Gamma) = 
\frac{1}{2\pi} \oint_\Gamma \frac{ v_x dv_y - v_y dv_x }{ v_x^2 + v_y^2 }
 \in \mathbb Z \,.
 \la{index1}
\end{equation}
Here $\Gamma$ is a simple closed curve on the ($x,y$)-plane that does not pass through 
any fixed point of $\vv(\x)$. 
For a curve that encircles a single isolated  center $p$ once in the counterclockwise 
direction, the winding number  is $i_\vv(p;\Gamma) = +1$ independently of the sign of 
(\ref{winding}). For a source and a sink, the winding number is also one. 
A saddle is the only fixed point with a negative winding number. For a single 
isolated saddle the winding number is $i_\vv(p;\Gamma) = -1$; see Appendix.

\subsection{Poincar\'e index formula}
\label{Sec:Poincare}

All the fixed points of a vector field $\vv (\x)$ can be detected by the Poincar\'e 
index formula. It states that for any simple closed curve $\Gamma$ that encloses 
finitely many fixed points $p_1, p_2,\cdots,p_k$ of $\vv (\x)$ in a counterclockwise 
direction, the sum of their winding numbers equals the following index 
\cite{Hartman-1964,Llibre-1998,Lloyd-1978}
\begin{equation}
\sum\limits_{j=1}^k i_{\vv}(p_j;\Gamma) = {\rm Index}(\Gamma) \ \equiv  \  
\mathcal X_\Gamma + \frac{1}{2} \left( I_\Gamma - E_\Gamma \right)	\,.
\la{index2}
\end{equation}
Here $\mathcal X_\Gamma$ is the Euler characteristic of the area that is bounded by 
the curve $\Gamma$, and $I_\Gamma$ is the number of concave curve segments with tangencies 
that are internal to the area that is bounded by $\Gamma$, and $E_\Gamma$ is the number of 
convex curve segments where the tangencies are external to  $\Gamma$; see Figure 
\ref{Fig:Fixed-points}.  The Poincar\'e index formula is a generalization of the 
Poincar\'e-Hopf index theorem \cite{Milnor-1965}. The latter assumes,  that the vector of 
$\vv (\x)$ always points towards the outward (or inward) normal direction along the 
boundary curve of a region. We remind that for any disk the Euler characteristic is 
$ \mathcal X_\Gamma =1$.

Besides the superflow velocity $\vv(\x,t)$ we introduce another vector field, that 
depends only on the modulus of the complex wave function $\psi(\x,t)$:  
\begin{equation}
\vw(\x, t) = \Grad |\psi(\x,t)| \,.
\label{Eq:W}
\end{equation}
Unlike the phase, the modulus of a normalizable wave function is a smooth, single valued and 
strictly non-negative function that vanishes both  at $\x\to\infty$ and at vortex cores. 
Consider any simple closed trajectory around the origin, with a sufficiently large radius 
to encircle all the critical points of $|\psi(\x,t)|$ but so that the vector field 
\Eqref{Eq:W} on the trajectory does not vanish.  
Along this trajectory the vector field \Eqref{Eq:W} always points towards its interior, 
so that the winding number \Eqref{index1} has the value $i_\vw=+1$, independently of the 
number of vortex cores that are encircled (note that the circulation \Eqref{index1} of 
this vector field vanishes). A single vortex core is an isolated minimum value critical 
point of $|\psi(\x,t)|$, thus it is a source of the vector field \Eqref{Eq:W} 
and as such it contributes $i_\vw=+1$ to the index \Eqref{index2}. Since a stagnation point
is the only fixed point of a vector field with a negative winding number $i_\vw=-1$, 
each vortex core must always be balanced by an accompanying stagnation point of $|\psi(\x,t)|$.  

Note that for certain vortex configurations there can be degeneracies in the fixed point 
structure of \Eqref{Eq:W}. This is the case {\it e.g.} when a single vortex is located at 
the center of the trap, and the vortex core is encircled by a nodal line where the modulus 
$|\psi(\x,t)|$ has a maximum value. The vector field \Eqref{Eq:W} then has a degeneracy 
circle and instead of \Eqref{index1} the appropriate index theoretic analysis is based on 
the more general concept of a degeneracy index introduced in \cite{Ruan-2019}.

\subsection{Gross-Pitaevskii equation}
\label{Sec:Gross-Pitaevskii}

We proceed to apply the Poincar\'e index formula \Eqref{index2} to characterize the 
minimal energy configurations of a trapped Bose-Einstein condensate of $N$ alkali atoms, 
with specified values of the particle number and of the macroscopic angular momentum.  
The trap is axially symmetric and non-rotating, and without loss of generality, we choose 
the trapping potential to be harmonic $V(\x) = |\x|^2/2$. The set-up can model an 
anisotropic three dimensional trap, resulting in an oblate, essentially 
two dimensional spheroid condensate $\mathcal D$.
The atoms interact with each other via a repulsive short range pair potential, 
so that in the limit where $N$ becomes large, the condensate 
can be modelled by a solution of the time-dependent two dimensional Gross-Pitaevskii 
equation. This is a nonlinear Schr\"odinger equation for the complex valued wave 
function $\psi(\x,t)$ that describes  the condensate as a coherent macroscopic 
quantum state. 

With an appropriate choice of various scales and parameters, the relevant 
two dimensional time-dependent Gross-Pitaevskii equation reads as follows 
\cite{Bao-2013}:
\begin{equation}
i\partial_t\psi=-\frac{1}{2}\Grad^2\psi + \frac{|\x|^2}{2}  \psi+g|\psi|^2\psi 
\equiv \ \frac{\delta F} {\delta \psi^\star} \,.
\la{GPeq}
\end{equation}
The dimensionless coupling $g$ specifies the strength of the pairwise interatomic 
interactions. In a typical experiment with $10^4$--$10^6$ atoms its numerical values 
are $g\sim10^1$--$ 10^3$. For a detailed discussion of the conventions used here, 
see \cite{Garaud-2021}.

We are interested in the ground state solution of \Eqref{GPeq} that minimizes the  
Gross-Pitaveskii free energy
\begin{equation}
F=\int d^2x \left\{\frac{1}{2}|\Grad\psi|^2 + \frac{|\mathbf x|^2}{2}   |\psi|^2 
+\frac{g}{2}|\psi|^4\right\}\,.
\la{F}
\end{equation}
Since $F$ is a strictly convex functional, its only critical point is the absolute 
minimum $\psi(\x) \equiv 0$.  But in this case there is no cold atom condensate in 
the trap. Thus, in the presence of any cold atom condensate, the free energy can 
not have any critical point. As a consequence it follows that, as detailed in 
\cite{Alekseev-2020}, the ground state solution of the Gross-Pitaevskii equation can 
be considered "time-crystalline" \cite{Wilczek-2012} provided there are conserved 
quantities:
Besides the free energy $F$ the time evolution \Eqref{GPeq} conserves two additional 
quantities, as Noether charges. One of these is   
\begin{equation}
\left\langle\hat {\mathrm N}\right\rangle \ \equiv  \ \int d^2 x \, \psi^\star \psi  
\ \equiv \ N = 1 \,,
\la{N}
\end{equation}
that counts the number of atoms. For clarity we have chosen the parameter values in  
\Eqref{GPeq}, \Eqref{F} so that the numerical value \Eqref{N} is normalized to $N=1$. 
For an axially symmetric trap, the macroscopic angular momentum along the $z$-axis 
\begin{equation}
\left\langle\hat {\,\mathrm {L}}_z\right\rangle  \ \equiv \ \int d^2x\, 
\psi^\star(- i \ez \cdot \x\wedge \Grad)\psi \ \equiv \ L_z = l_z \,,
\la{Lz}
\end{equation}
is also conserved. The numerical value $l_z$ of the canonical angular momentum of 
the condensate
\begin{equation}
\hat {\,\mathrm {L}}_z = - i \ez \cdot \x\wedge \Grad \equiv - i \partial_\theta \,,
\la{Lzop}
\end{equation} 
with  $\theta$ the polar angle around the $z$-axis, is a free parameter that can take an
arbitrary value; see  \cite{Garaud-2021}.

The  {\it r.h.s.} of \Eqref{GPeq} can not vanish unless $\psi \equiv 0$, so that for 
non-vanishing values of the Noether charges, the minimum of $F$ can not be a critical 
point of $F$. Thus, to construct the ground state wave function of \Eqref{GPeq}, we use 
methods of constrained optimization and minimize the free energy \Eqref{F} at the 
{\it fixed} values of the Noether charges \Eqref{N} and \Eqref{Lz}. 
The Lagrange multiplier theorem \cite{Marsden-1999} states that the minimum of \Eqref{F} 
can be found as a critical point of
\begin{equation}
F_\lambda=F + \lambda_N  (N-1)+  \lambda_z ( L_z - l_z)    \,,
\la{Fl} 
\end{equation}
where the Lagrange multipliers $\lambda_N$, $\lambda_z$ respectively enforce the values 
$N=1$ and $L_z=l_z$ of the Noether charges. The critical points of $F_\lambda$ obey
\begin{equation}
-\frac{1}{2}\nabla^2 \psi 
+ \frac{|\x|^2}{2} \psi + g |\psi|^2 \psi 
= - \lambda_N \psi + \lambda_z ( i \ez \cdot \x\wedge \Grad )  \psi ,
\la{Flambda1}
\end{equation}
together with the two conditions \Eqref{N} and \Eqref{Lz}. From these three equations we 
solve for the critical point wave function $\psi_{cr}(\x)$ and for the ensuing Lagrange 
multiplier values  $\lambda^{cr}_{N},  \lambda^{cr}_{z}$.
Note that both Lagrange multipliers are time independent \cite{Alekseev-2020}, and that 
they cannot vanish simultaneously.

We focus on the critical points of \Eqref{Fl} that are also minima of the free energy 
\Eqref{F}. Let $\{\psi_{min}(\x), \lambda^{min}_{N}, \lambda^{min}_{z}\}$ denote such 
a configuration. If $\psi_{min}(\x)$ is an initial value of the Gross-Pitaevskii equation 
\Eqref{GPeq}, then it obeys the {\it linear} time evolution  
\begin{equation}
 i \partial_t \psi = -\lambda^{min}_{N}\psi  
			 		+ i\lambda^{min}_{z} \ez\cdot\x \wedge\Grad\psi  \,.
\la{Heq} 
\end{equation}

We recall \Eqref{Lzop}, denote $\sigma=\lambda^{min}_{N}/\lambda^{min}_{z}$, and define
\[
A_\theta = \mathbf e_z \cdot \mathbf x \wedge \Grad \tan^{-1} (x/y)
\]
This is a vector field on the plane with a center at the origin ($x,y$)=(0,0). 
The winding number \Eqref{index1} is $i_{\mathbf A} = +1$  for any trajectory 
that encircles the origin once in counterclockwise direction; we note that $A_\theta$ 
is akin the azimuthal component of the vector potential of a line vortex along the 
$z$-axis.  The time evolution \Eqref{Heq} can further be written as
\begin{equation}
 i \partial_t \psi =-\lambda^{min}_{z}( \hat{\,\mathrm L}_z + \sigma\, A_\theta ) \psi 
 \ \equiv \ 
-\lambda^{min}_{z} {\hat{\,\mathrm L}_z}^{cov}  \psi \,.
\label{Heq1} 
\end{equation}

In the presence of $A_\theta$ the rotations around the $z$-axis {\it i.e.} changes in 
the polar angle $\theta$ are generated by the covariant angular momentum operator 
${\hat{\,\mathrm L}_z}^{cov}$ instead of the canonical angular momentum \Eqref{Lzop}. 
Thus the equation \Eqref{Heq} describes the rotation of the initial wave function 
$\psi_{min}(\x)$ around the center of the trap, with angular velocity $\lambda^{min}_{z}$.

In the following we are interested in the consequences of the Poincar\'e index formula  
\Eqref{index1}, in the case of a vector field $\vv(\x,t)$ that describes the velocity 
of the superflow \Eqref{v} of the minimal energy solution of the Gross-Pitaevskii 
equation \Eqref{GPeq}) with initial condition $\psi(\x, t=0) =  \psi_{min}(\x)$. 
We focus on the topological and geometrical properties of its fixed point structure, 
as the value of the angular momentum $l_z$ is changed. For this we analyze the profile 
of the superflow vector field, by employing the Poincar\'e index formula in combination with 
various different trajectories $\Gamma$.

\section{Numerical analysis}
\label{Sec:Results}

We have numerically constructed the critical points of \Eqref{Fl} that minimize the 
Gross-Pitaevskii free energy \Eqref{F}. For clarity we choose positive $l_z >0$ and with 
$g=$ 5, 100 and 400. Since the topological properties can not change with continuous 
changes of $g$, the results are displayed only for $g=400$. (For detailed results with 
other values of $g$, see \cite{Garaud-2021}). The problem is discretized within a 
finite-element framework \cite{Hecht-2012}, and the constrained optimization problem 
is solved using the Augmented Lagrangian Method; numerical methods are discussed in 
details in \cite{Garaud-2021}.
Whenever $l_z\not=0$ the minimum free energy solution is a configuration with vortices 
and superflow stagnation points; the former are centers of the vector field $\vv(\x)$,
and the latter are saddles of $\vv(\x)$. The ensuing time-evolution \Eqref{GPeq}, 
\Eqref{Heq} is timecrystalline, as  it always depends on the time variable $t$ in a 
nontrivial fashion \cite{Garaud-2021}. 
%

\begin{figure*}[!tb]
\hbox to \linewidth{ \hss
\includegraphics[width=0.99\linewidth]{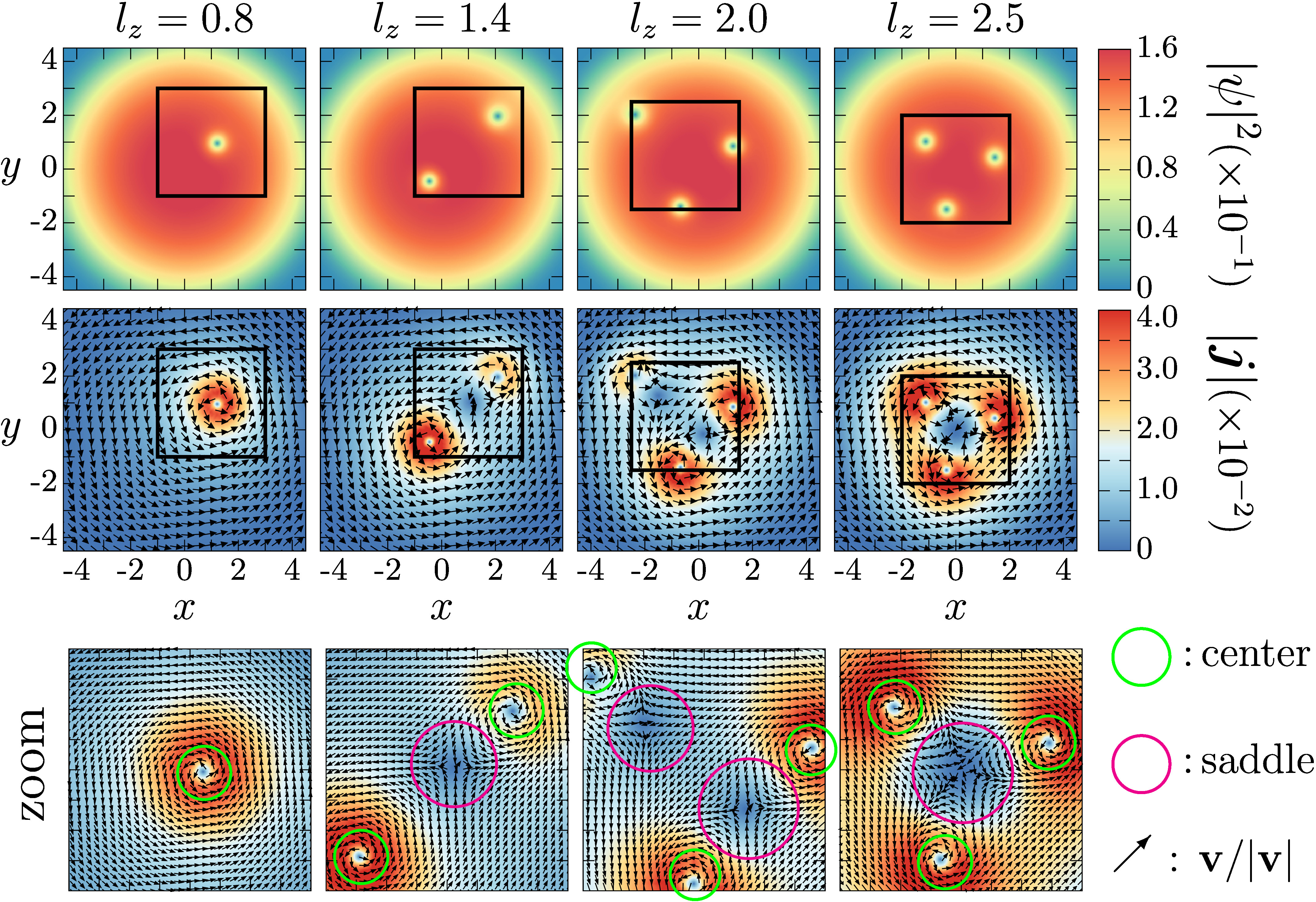}
\hss} 
\caption{ 
The panels on the different columns show examples of the minimal energy configurations 
of \Eqref{F}, that satisfy $L_z=l_z$ and $N=1$ for the dimensionless coupling $g=400$. 
The panels on the top row show the density $|\psi|^2$; it vanishes at the vortex core. 
The squares denote the regions that are zoomed-in, on the bottom panels. The middle panels 
show the corresponding vector field $\vv(\x)$. The color map shows the amplitude of the 
current $\vj=|\psi|\vv$, while the arrows show the superflow vector field  $\vv(\x)$. 
The bottom row shows the corresponding data zoomed closer to the various fixed points 
of $\vv(\x)$. 
The green and magenta circles highlight the vortex cores {\it i.e.} centers of  $\vv(\x)$
and stagnation  points {\it i.e.} saddles
of  $\vv(\x)$,  corresponding to different trajectories $\Gamma$ in \Eqref{index2}.
Both the vortex cores and the stagnation points are characterized by a vanishing superflow 
$|\vj (\x)|=0$. Unlike the vortex cores, where the density $|\psi|$ vanishes, at the 
stagnation points $|\psi|\neq0$. This can be seen by comparing the data for the current 
$\vj:=|\psi|\vv$, with the density $|\psi|^2$ data. 
}
\label{Fig:Vortices:1}
\end{figure*}

The number of vortices increases with $l_z$. Each new vortex core enters the condensate 
at the boundary of the disk $\mathcal D$, and moves towards the trap center as $l_z$ 
increases. Since the Euler characteristic of a disk is  $\mathcal X_\Gamma =1$, the 
Poincar\'e index formula ensures that, on the entire trapping disk $\mathcal D$, there 
is always one more vortex core  {\it i.e.} center than there are stagnation points {\it i.e.} saddles. 
That is, when the closed 
curve $\Gamma$ coincides with the perimeter of the entire disk $\mathcal D$ in 
counterclockwise direction, the winding number  \Eqref{index1} always has the value 
$i_{\mathbf v}=1$.

\subsection{Small angular momentum}
\label{Sec:Simulations} 

The Figure \ref{Fig:Vortices:1} shows examples of vortices and stagnation points for 
different values of the macroscopic angular momentum $l_z$. The corresponding solutions 
of Eq.~\Eqref{GPeq}, describe how these structures rotate uniformly around the symmetry 
axis, with constant angular velocity $-\lambda^{min}_{z}$; see Eq.~\Eqref{Heq1}.

\begin{figure*}[!tb]
\hbox to \linewidth{ \hss
\includegraphics[width=\linewidth]{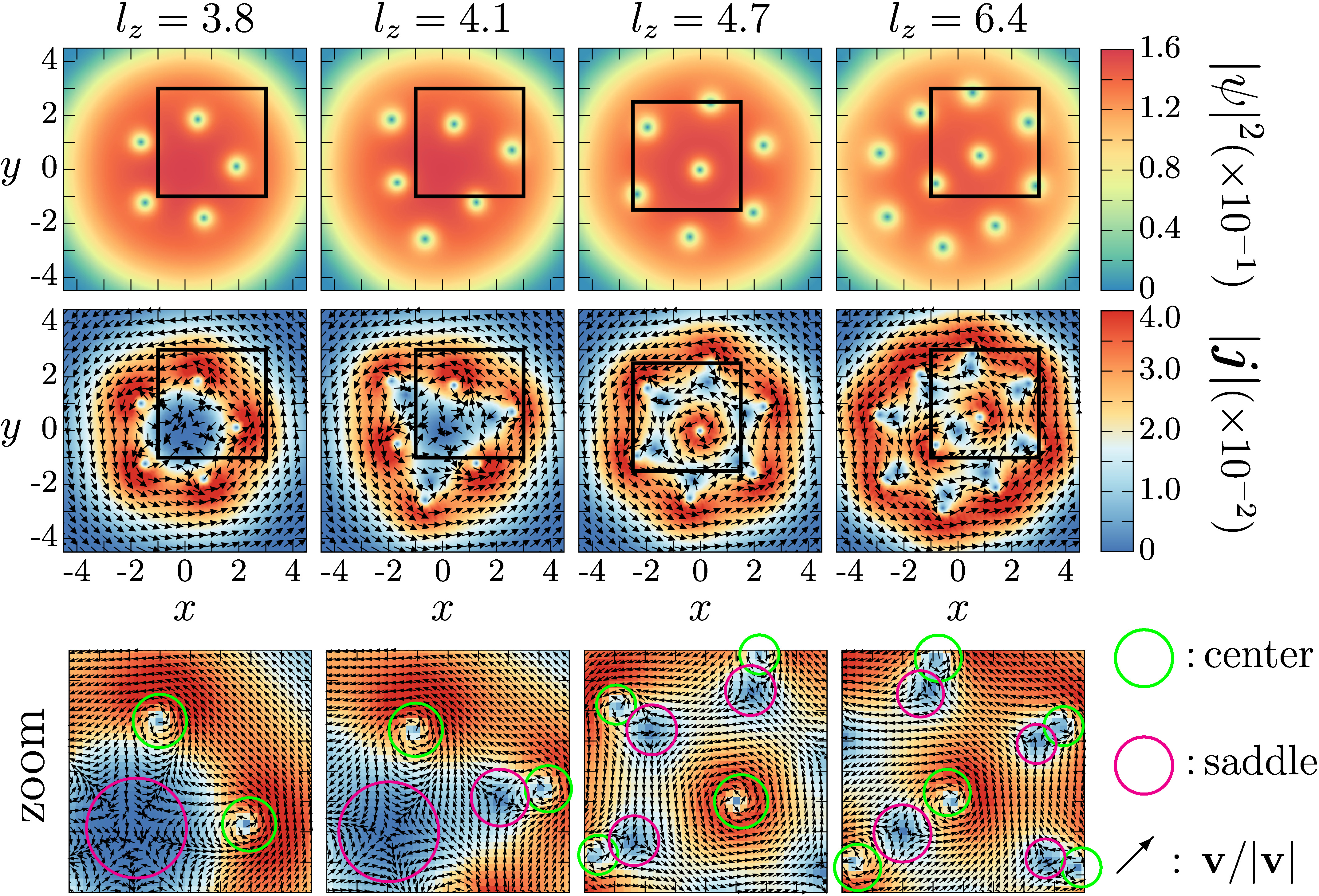}
\hss} 
\caption{ 
The panels show the minimal energy configurations for larger values of the angular 
momentum $L_z=l_z$ and $N=1$ for the dimensionless coupling $g=400$. The displayed 
quantities are the same as in  \Figref{Fig:Vortices:1}.
According to the Poincar\'e index formula there are always one less stagnation points 
than there are vortex cores. The formation of paired vortex cores and stagnation points 
is in resemblance of the Kosterlitz-Thouless transition that occurs {\it e.g.} in the 
2D XY model and the planar rotor model.
}
\label{Fig:Vortices:2}
\end{figure*}

For $0<l_z<1$ the condensate features a unique vortex  core that is located off the trap 
center. This is the regime $l_z=0.8$ of \Figref{Fig:Vortices:1}, which features 
a single eccentric vortex core as the unique fixed point of $\vv(\x)$.
Next, when $l_z=1.4$, the minimal energy wave function features two additional 
fixed points of $\vv(\x)$, a vortex core and a stagnation point. Upon increasing $l_z$, 
the second vortex core together with the stagnation point move toward the trap center, 
until they form a symmetric pair of vortex cores with a single stagnation point in between 
(not shown). With a further increase of $l_z$, a third vortex core and a second stagnation 
point of $\vv$ now appear, as can be seen when $l_z=2.0$.
Finally, for $l_z=2.5$ the third vortex core enters closer to the center of the condensate, 
to form a symmetric triangle. This is accompanied by a merging of the two stagnation points  
into a higher degree saddle; there is a degenerate saddle with winding number 
$i_{\mathbf v}=-2$ at the center of the disk and it is surrounded by the three vortex cores 
as centers.

\subsection{A Kosterlitz-Thouless analogy}

The Figure \ref{Fig:Vortices:2} shows the evolution of the minimum energy state, 
when $l_z$ is further increased: We observe a remarkable transition in the geometric 
pattern of vortex cores  and stagnation points; the topology closely resembles the geometry 
of vortices and antivortices in the standard presentation of the Kosterlitz-Thouless 
transition in the case of the two dimensional XY-model; see details in the Appendix. 

First, when  $l_z = 3.8$, there are five single vortex cores that surround an aggregated 
stagnation point structure, with winding number $i_\vv=-4$: The overall geometric pattern 
remains the same as in Figure \ref{Fig:Vortices:1}. 

But when $l_z$ increases further, there is a transition and instead of individual isolated 
vortex cores  and an aggregation of stagnation points, we start observing a formation of paired 
vortex cores  and stagnation points: In panels $l_z=4.1$ of Figure \ref{Fig:Vortices:2} the total 
number of vortices is  six. There are now three individual vortex cores, while three vortex cores 
have each paired with a stagation point with winding number $i_\vv=-1$.  According to the 
Poincar\'e index formula the individual stagnation point at the center of the condensate then 
has winding number $i_\vv=-2$. 
The configuration for $l_z=4.7$ in \Figref{Fig:Vortices:2} shows how the transition proceeds:
The configuration consists of seven vortex cores, arranged in an almost triangular lattice. 
The central vortex core here is isolated, while the six outer vortex cores with $i_\vv=1$ are tightly 
bound to six  stagnation points, each with a winding number $i_\vv=-1$.
Finally the configuration with $l_z=6.4$ consists of a pair of vortex cores that are bound to a 
central stagnation point,  surrounded by eight bound pairs of vortex cores and 
stagnation points. 

\begin{figure}[!tb]
\hbox to \linewidth{ \hss
\includegraphics[width=0.6\linewidth]{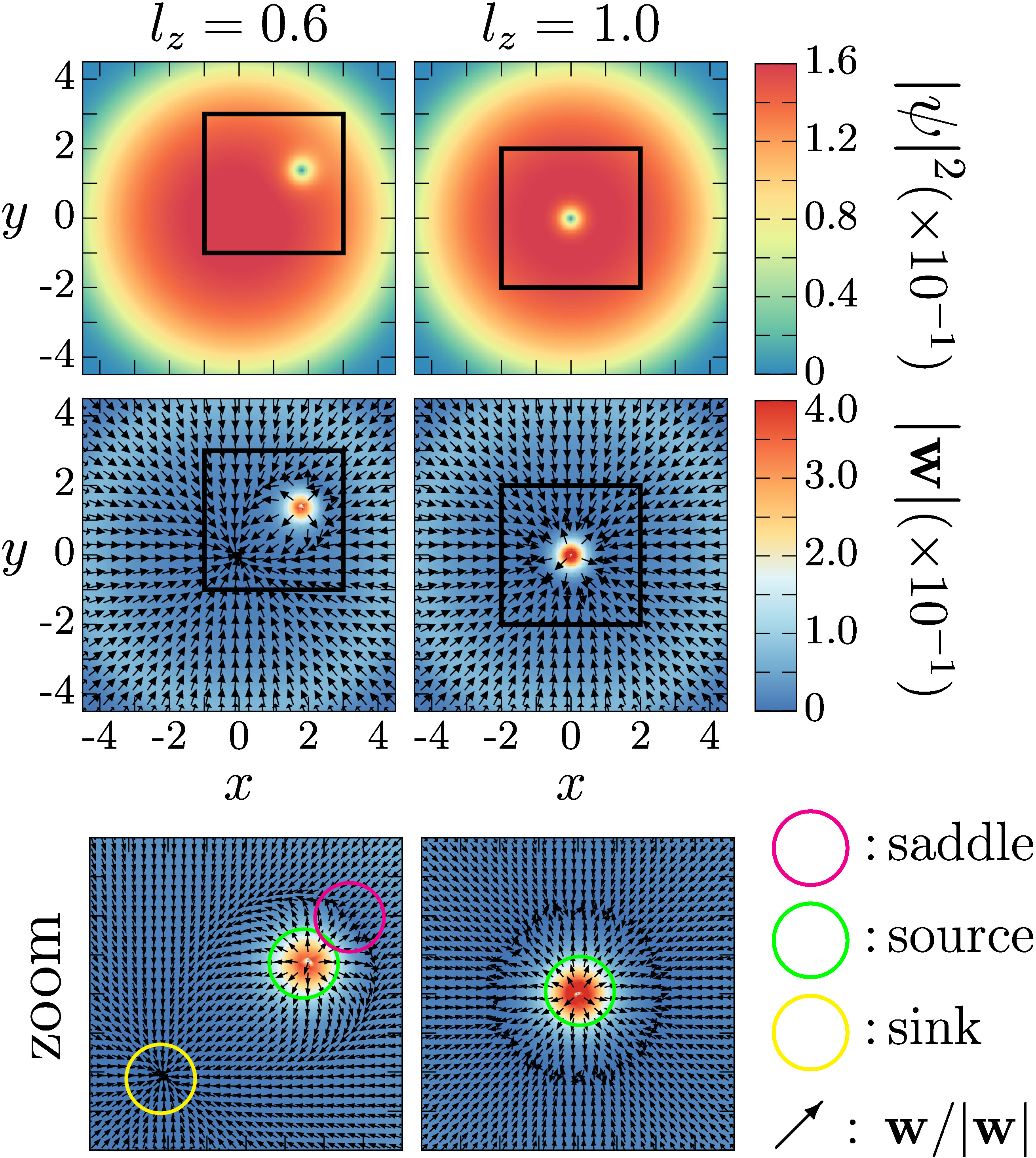}
\hss} 
\caption{ 
The panels show the minimal energy configurations for values of the angular 
momentum $l_z=0.6$ and $1$ and $N=1$ for the dimensionless coupling $g=400$. The 
panels on the top row display the density $|\psi|^2$.
The middle panels display the corresponding vector field $\vw(\x)$, and the bottom 
line shows the corresponding data zoomed closer to the fixed points of $\vw(\x)$. 
The green and yellow circles respectively highlight the source (the vortex) and sink of 
$\vw(\x)$. The saddle of $\vw(\x)$ is represented by magenta circle. All these 
circles correspond to different trajectories $\Gamma$ in \Eqref{index2}, for the vector 
field $\vw(\x)$.
}
\label{Fig:Vortices:3}
\end{figure}

When the value of $l_z$ is further increased, we observe that the pattern identified 
in Figures \ref{Fig:Vortices:2}  are repeated: The difference 
between the vortex cores and stagnation points is always one. When $l_z$ increases the new 
vortex cores  and stagnation points enter the disk together, as tightly bound pairs, and the pairs 
form structures that are akin co-centric Abrikosov lattices. With the additional structure, 
that as dictated by the Poincar\'e index formula there is commonly either a single vortex core, 
or a triplet of two vortex cores  and a single stagnation point near  the center of the trap.

\subsection{Degeneracy circle}

Finally, the figure \ref{Fig:Vortices:3} summarizes the main features that can be 
observed for the vector field $\vw(\x)$ \Eqref{Eq:W}, that is associated with density 
gradients, in the case of a single vortex:  The configuration $l_z=0.6$ in the Figure 
features a single eccentric vortex core as a source, and a sink of $\vw(\x)$. The Poincar\'e 
index formula dictates that there is also a saddle. In the case where $l_z=1$, there is a nodal 
line encircling the source. This is a degeneracy circle with winding number $i_\vw=-1$ 
\cite{Ruan-2019}.

\section*{Conclusion}
\vskip 0.3cm

We have used the Poincar\'e index formula in combination with numerical simulations, 
to study the topology and geometry of solutions to the two dimensional Gross-Pitaevskii 
equation at different angular momentum values. 
The minimum energy solution of the equation models the ground state of a 
cold atom Bose-Einstein condensate in an axially symmetric disk-like, non-rotating 
harmonic trap. 
We have varied the angular momentum that is supported by the ground state wave function, 
and confirmed that the difference in the number of vortex cores and in the number of stagnation
points is always equal to one. In particular we have observed how the vortex cores and the 
stagnation points enter the condensate, always in pairs, when the angular momentum increases. 
We have observed how the vortex cores repel each other to form co-centric Abrikosov 
lattices, while for small values of the angular momentum the individual stagnation points tend to 
aggregate into higher degree stagnation ipoints at the trap center. But when the angular 
momentum increases, there is a transition and the stagnation points start to pair up with 
vortex cores. The ensuing structures always rotate around the trap center at a constant 
angular velocity. 

This transition that occurs when the angular momentum increases has a geometric pattern 
that closely resembles the  geometric pattern of the Kosterlitz-Thouless transition in 
{\it e.g.} two dimensional XY-model; see Appendix.  But note that in the case of the standard 
XY-model, the energy of the vortices and antivortices diverges when the cutoff scale for the core 
size vanishes. In the present case the energy of the individual vortices and stagnation points 
is always finite but increases logarithmically in the length scale that characterizes the 
range of the confining trap potential. Furthermore, while in the constant density 
approximation the core contribution to the energy of vortices diverges logarithmically,
in the case of stagnation points their core contribution to energy remains finite even in 
this limit.
Moreover, instead of temperature the external control parameter is the angular momentum 
$l_z$, such that at large value of $l_z$ the geometrical pattern resembles that of 
small temperature XY-model, and for small  values of $l_z$ the topological charge 
structure resembles that of the XY-model high temperature phase. It remains 
to be clarified whether the present transition is truly a topological phase 
transition \cite{Frohlich-1981}.

\begin{acknowledgments}
We thank Li You for a valuable comment. AJN also thanks S. Komineas, S. Morampudi 
and F. Wilczek for comments. This work is supported by the Swedish Research Council 
under Contract No. 2018-04411. The research by AJN is partially supported by the 
Carl Trygger Foundation Grant CTS 18:276. AJN also acknowledges collaboration under 
COST Action CA17139. Nordita is partially supported by Nordforsk. 
The computations were performed on resources provided by the Swedish National 
Infrastructure for Computing (SNIC) at National Supercomputer Center at Link\"{o}ping, 
Sweden. 

\end{acknowledgments}

\appendix
\setcounter{section}{0}
\setcounter{equation}{0}
\renewcommand{\theequation}{\Alph{section}\arabic{equation}}

\section{Fixed point structure of the velocity and of the XY-model}

\begin{table}[!htb]
	 \graphicspath{../Plots/fixed-points/}
  \centering
  \begin{tabular}{ | c | c | c |  }
    \hline
    {} & Vortex & Antivortex  \\ \hline

	\vspace{0.1cm}   
   	Phase $\varphi$ &	$-\arctan\frac{y}{x}$ &	$\arctan\frac{y}{x}$	
   	\\ \hline

 	\begin{minipage}{.15\linewidth} \vspace{0.1cm}
    	Phase $\varphi(x,y)$ 		\vspace{0.1cm}
	\includegraphics[width=0.75\linewidth,height=1.6\linewidth]{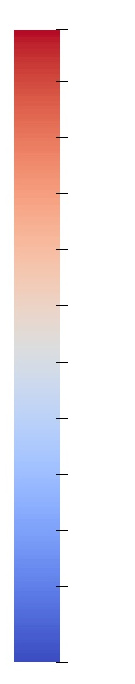}       
		\begin{picture}(0,0)%
    	\put(-22,0){\large $0$ }
	    \put(-24,92){\large $2\pi$ }
		\end{picture}
	\end{minipage}

    & \begin{minipage}{.35\linewidth}\vspace{0.1cm}
      \includegraphics[width=\linewidth]{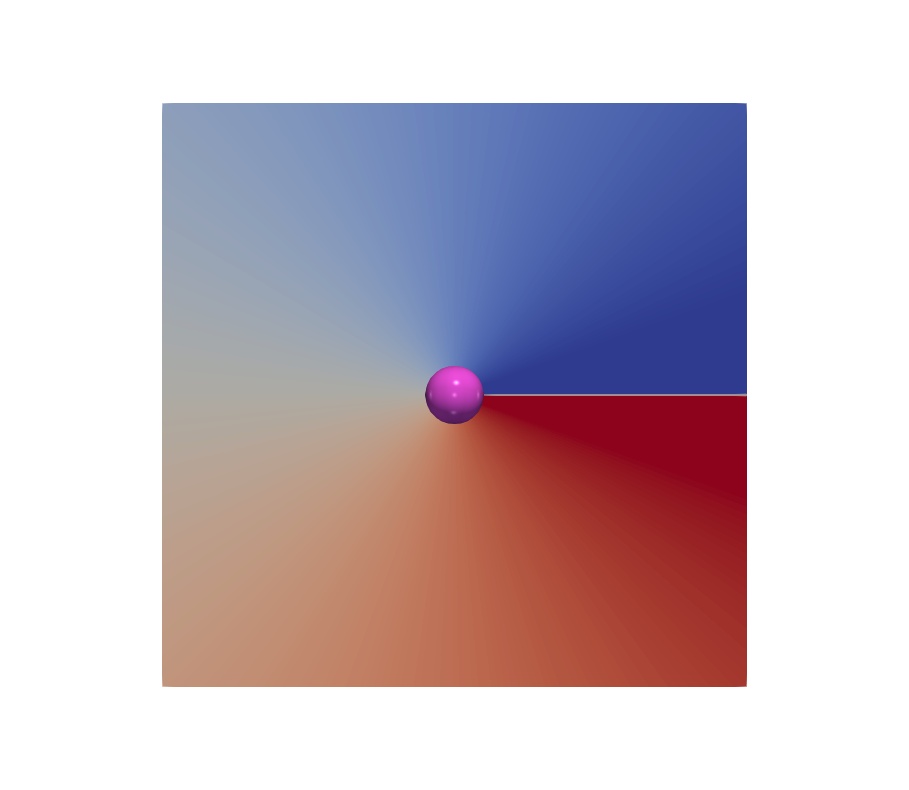}
    \end{minipage}
    & \begin{minipage}{.35\linewidth}\vspace{0.1cm}
      \includegraphics[width=\linewidth]{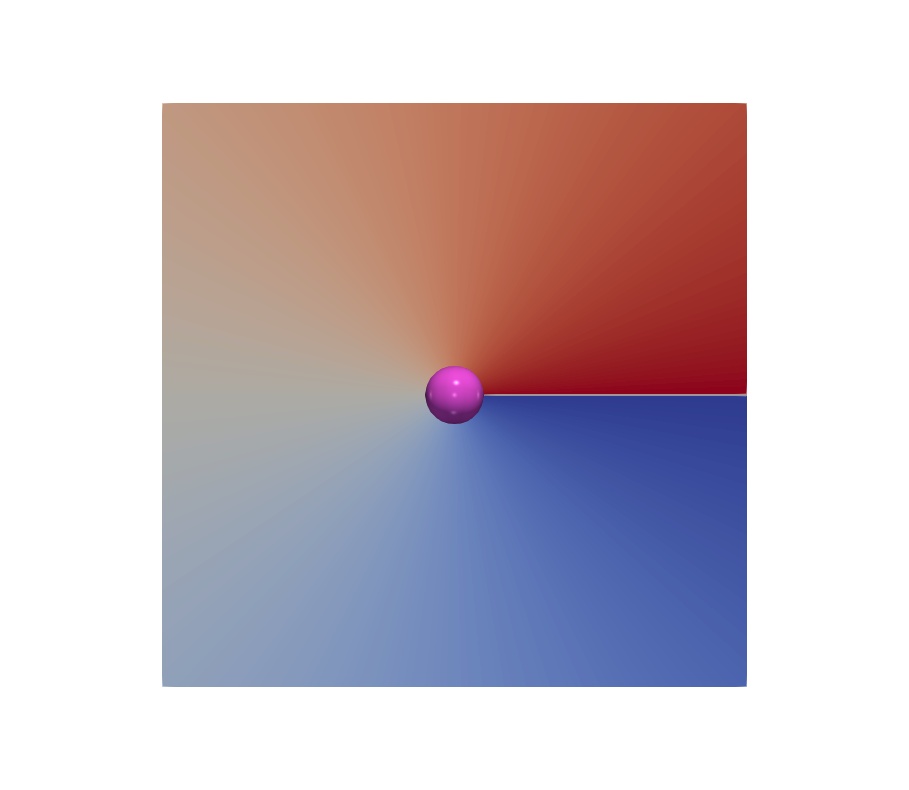}
    \end{minipage}
    \\ \hline

   \begin{minipage}{.185\linewidth}\vspace{0.2cm}  
   		XY-vector: 
   		$\vv_{\mathrm{XY}}$ 
    \end{minipage}
   	&	$\left(\frac{-y}{\sqrt{x^2+y^2}},\frac{ x}{\sqrt{x^2+y^2}}\right)$ 
  	&	$\left(\frac{-y}{\sqrt{x^2+y^2}},\frac{-x}{\sqrt{x^2+y^2}}\right)$ 
    \\ \hline
 
   \begin{minipage}{.185\linewidth}\vspace{0.2cm}  
   		XY-vector: 
   		$\vv_{\mathrm{XY}}$ 
    \end{minipage}
   & \begin{minipage}{.35\linewidth}\vspace{0.1cm}
      \includegraphics[width=\linewidth]{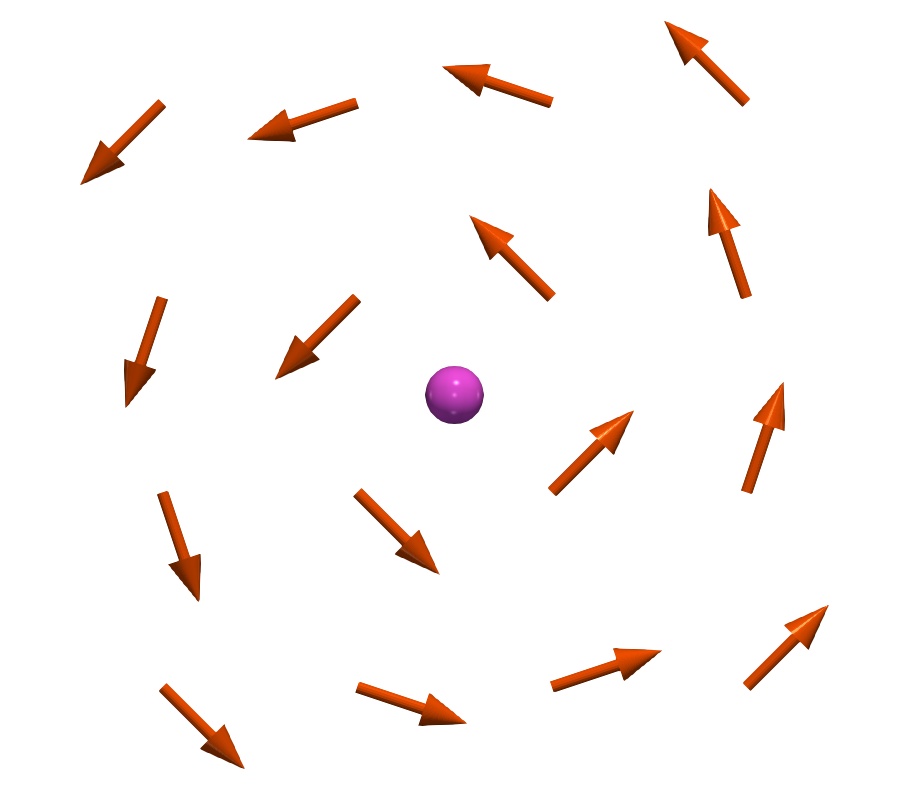}
   \end{minipage}
   & \begin{minipage}{.35\linewidth}\vspace{0.1cm}
      \includegraphics[width=\linewidth]{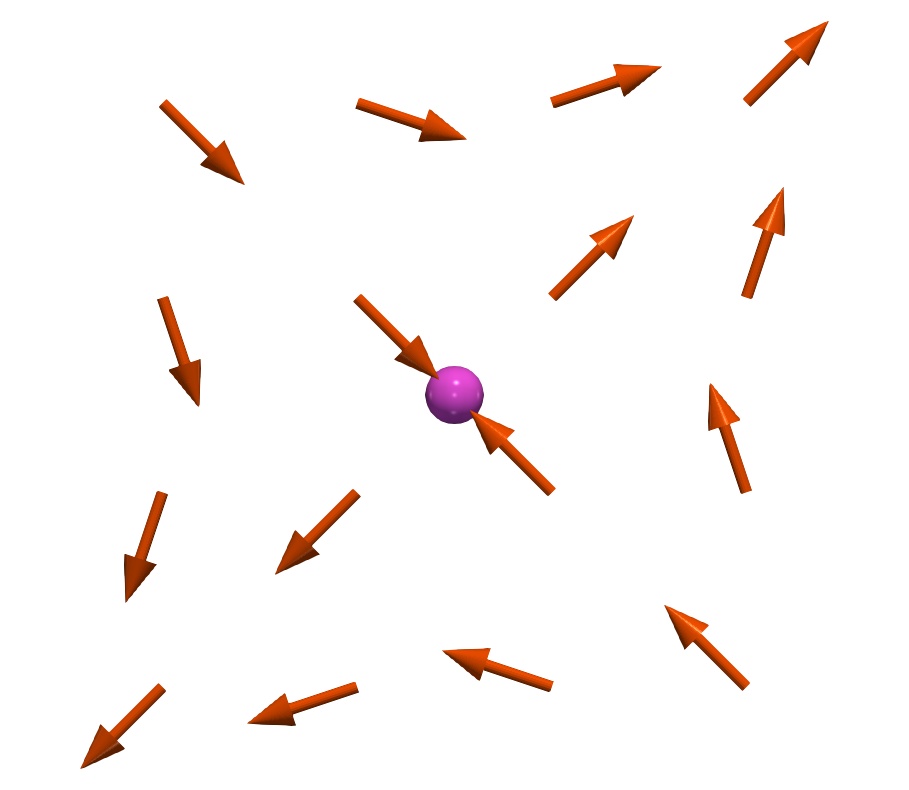}
   \end{minipage}
   \\ \hline

    \begin{minipage}{.1\linewidth}\vspace{0.1cm}
    	   Superflow  $\vv=\Grad\varphi$ \vspace{0.1cm}
	\end{minipage}
 	   &	$\left(\frac{-y}{x^2+y^2},\frac{x}{x^2+y^2}\right)$ 
  	   &	$\left(\frac{y}{x^2+y^2},\frac{-x}{x^2+y^2}\right)$ 
     \\ \hline
  
    \begin{minipage}{.1\linewidth}
 	   Superflow $\vv/|\vv|$ 
	    \end{minipage}
   & \begin{minipage}{.3\linewidth}	\vspace{0.1cm}
      \includegraphics[width=\linewidth]{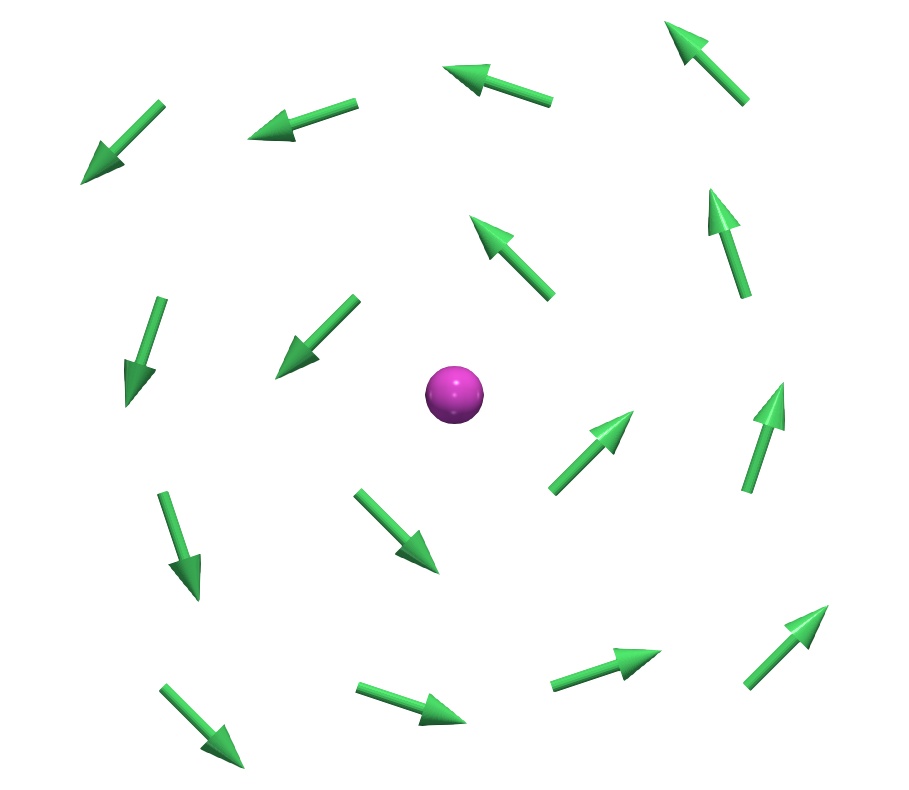}
    \end{minipage}

    &	   \begin{minipage}{.3\linewidth}\vspace{0.1cm}
      \includegraphics[width=\linewidth]{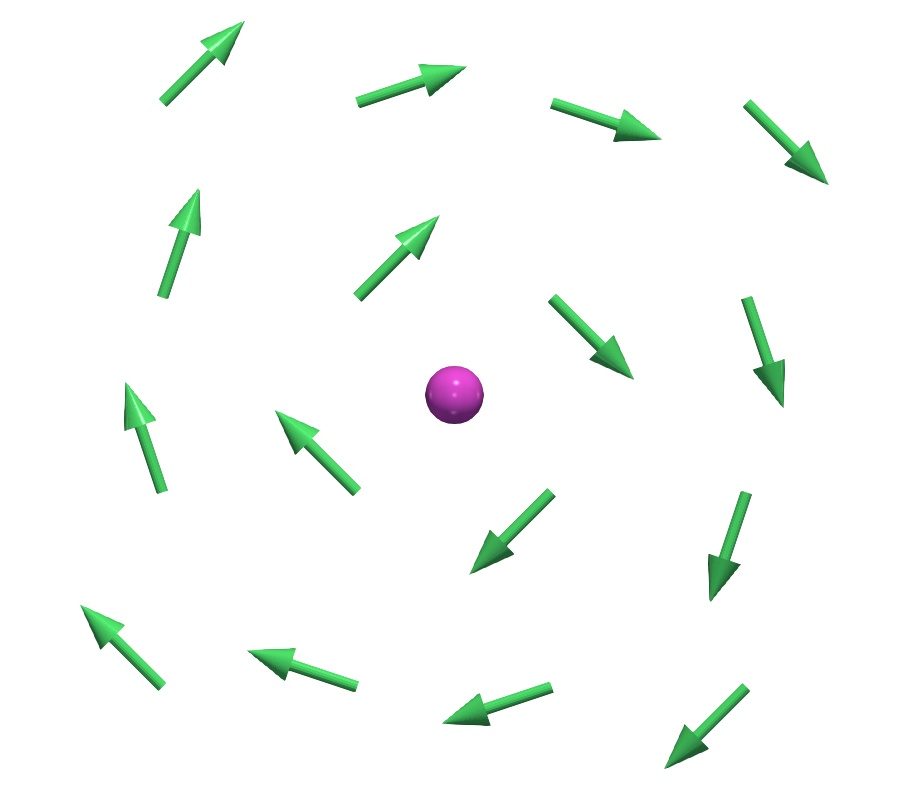}
    \end{minipage}
    
   \\ \hline
  \end{tabular}
  \caption{
	The Table compares the vector fields $\vv_{\mathrm{XY}}$ \Eqref{A-1} and $\vv$ 
	\Eqref{v} in the case of a vortex and an antivortex, corresponding to a fixed 
	point of the XY-vector $\vv_{\mathrm{XY}}$
	and the velocity vector field $\vv$ of the superflow. 
	The red dot is the phase singularity. 
	The vortex and the antivortex is respectively a center and a saddle, 
	of the XY-vector field $\vv_{\mathrm{XY}}$. On the other hand, they are both 
	centers of the vector 
	field $\vv$.
  }
  \label{Table:illustration:1}
\end{table}

Here, we compare in detail the fixed point structure as it is commonly presented in 
the XY-model \cite{Kosterlitz-1972,Kosterlitz-1973,Nobel},  and the fixed point 
structure of the superflow velocity that is presented in the main text.

As shown in the Table~\ref{Table:illustration:1}, a vortex and an antivortex are 
respectively a center and a saddle-point of the XY-vector field which is associated  
to the phase of the wavefunction as 
\begin{equation}
\vv_{\mathrm{XY}} \ = \  (\cos\varphi,\sin\varphi) 
\,,~~~\text{where}~~\varphi:=\arg[\psi]	\,.
\la{A-1}
\end{equation}
On the other hand, vortices and antivortices are both centers of the superflow vector 
field $\vv$ in \Eqref{v}. 
Note that the phase $\varphi$ is defined up to a constant value, and thus 
$\vv_{\mathrm{XY}}$ is defined up to global rotations. It follows that a counterclockwise
center in $\vv_{\mathrm{XY}}$ as shown in Table II can be continuously rotated to a 
source, a sink or a clockwise center. The velocity $\vv$ of the superflow \Eqref{v} is 
invariant under these rotations and thus vortices remain unambiguously centers of $\vv$. 
Animations showing the effect of global rotations on $\vv_{\mathrm{XY}}$ can be 
found as Supplementary Material.

\begin{table}[!htb]
	 \graphicspath{../Plots/fixed-points/}
  \centering
  \begin{tabular}{ | c | c | c |  }
    \hline
    {} & Vortex/Vortex pair & Vortex/Antivortex  pair \\ \hline
    
	\vspace{0.1cm}	Phase $\varphi$ 
	& $-\arctan\frac{y}{x-x_0}-\arctan\frac{y}{x+x_0}$
	& $-\arctan\frac{y}{x-x_0}+\arctan\frac{y}{x+x_0}$	
    \\ \hline

  	\begin{minipage}{.15\linewidth} \vspace{0.1cm}
    	Phase $\varphi(x,y)$ 		\vspace{0.1cm}
	\includegraphics[width=0.75\linewidth,height=1.6\linewidth]{figure-5a}       
		\begin{picture}(0,0)%
    	\put(-22,0){\large $0$ }
	    \put(-24,92){\large $2\pi$ }
		\end{picture}
	\end{minipage}
   	& \begin{minipage}{.35\linewidth}\vspace{0.1cm}
      \includegraphics[width=\linewidth]{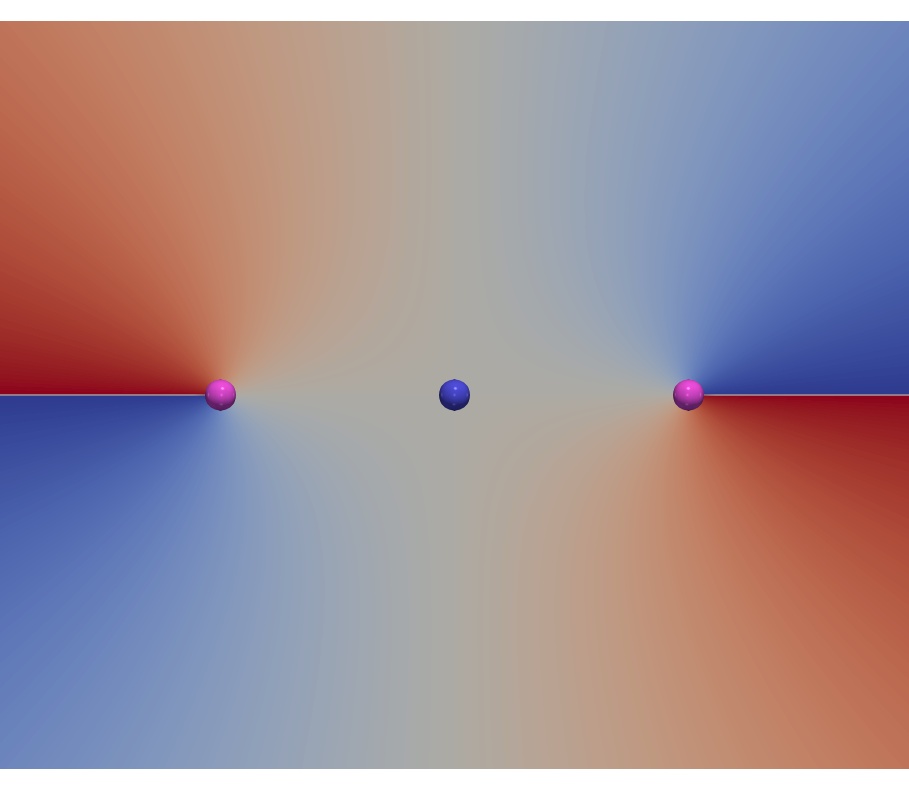}
	\end{minipage}

   	& \begin{minipage}{.35\linewidth}\vspace{0.1cm}
      \includegraphics[width=\linewidth]{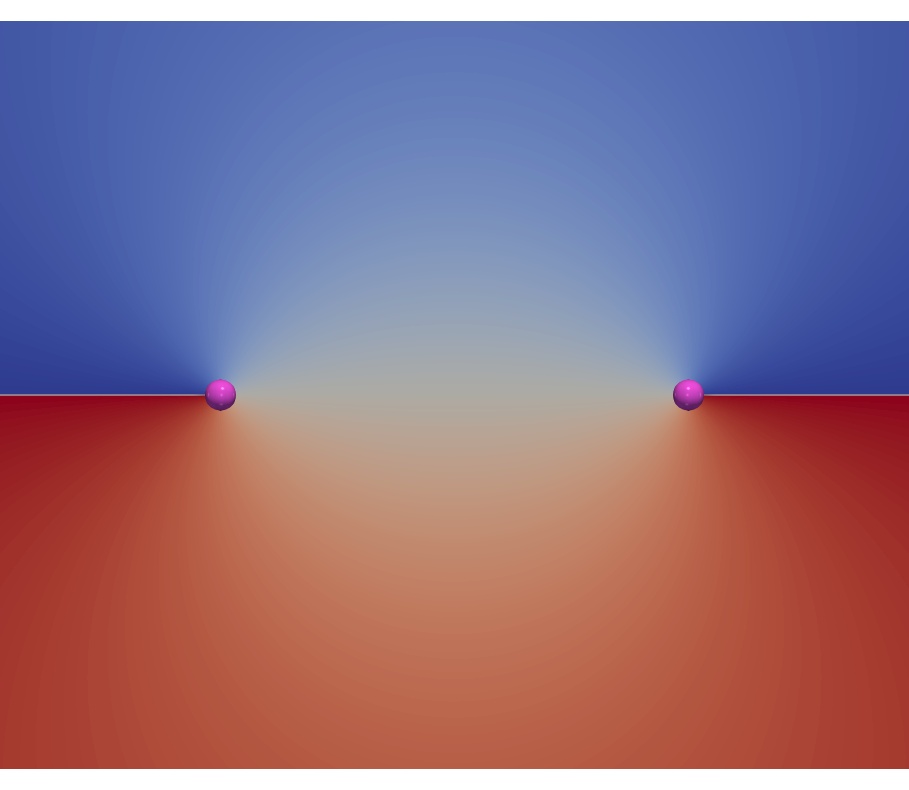}
   	\end{minipage}
    \\ \hline
 
   \begin{minipage}{.185\linewidth}\vspace{0.2cm}  
   		XY-vector: 
   		$\vv_{\mathrm{XY}}$ 
    \end{minipage}
   	& \begin{minipage}{.35\linewidth}\vspace{0.1cm}
      \includegraphics[width=\linewidth]{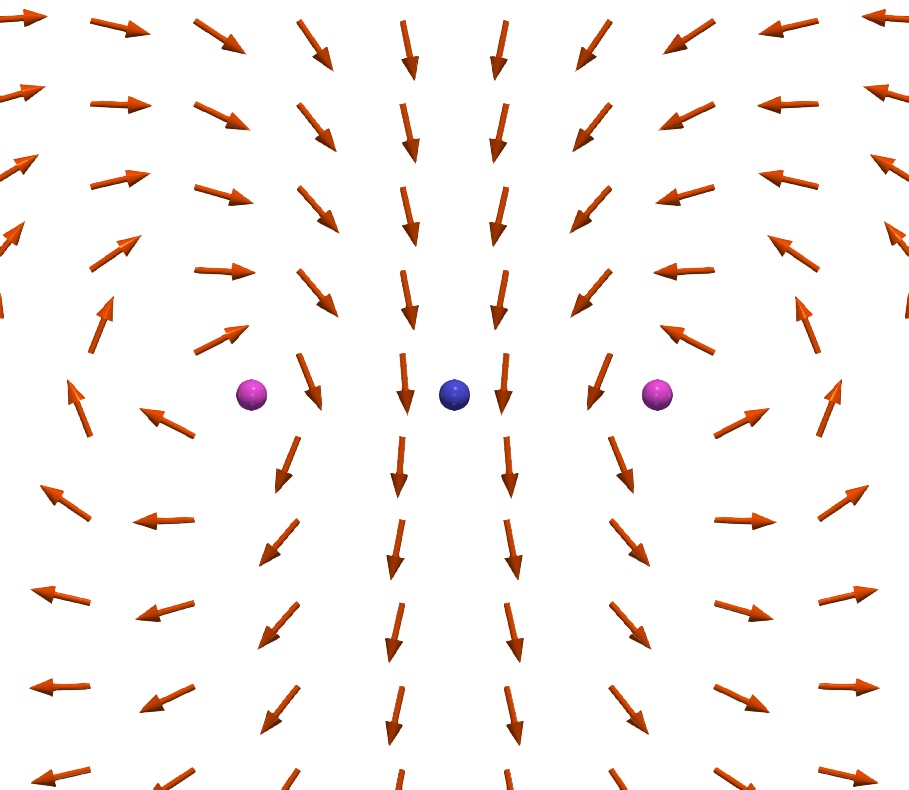}
    \end{minipage}
    & \begin{minipage}{.35\linewidth}\vspace{0.1cm}
      \includegraphics[width=\linewidth]{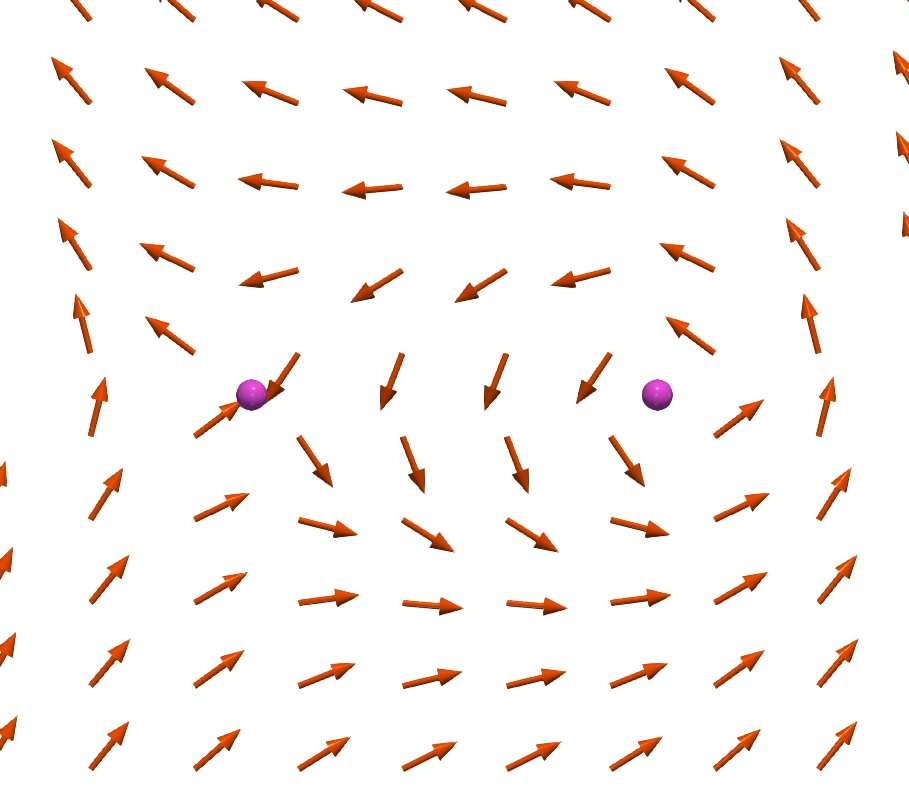}
    \end{minipage}
    \\ \hline
 
    \begin{minipage}{.1\linewidth}
 	   Superflow $\vv/|\vv|$, with $\vv=\Grad\varphi$
	\end{minipage}
   	& \begin{minipage}{.3\linewidth}	\vspace{0.1cm}
      \includegraphics[width=\linewidth]{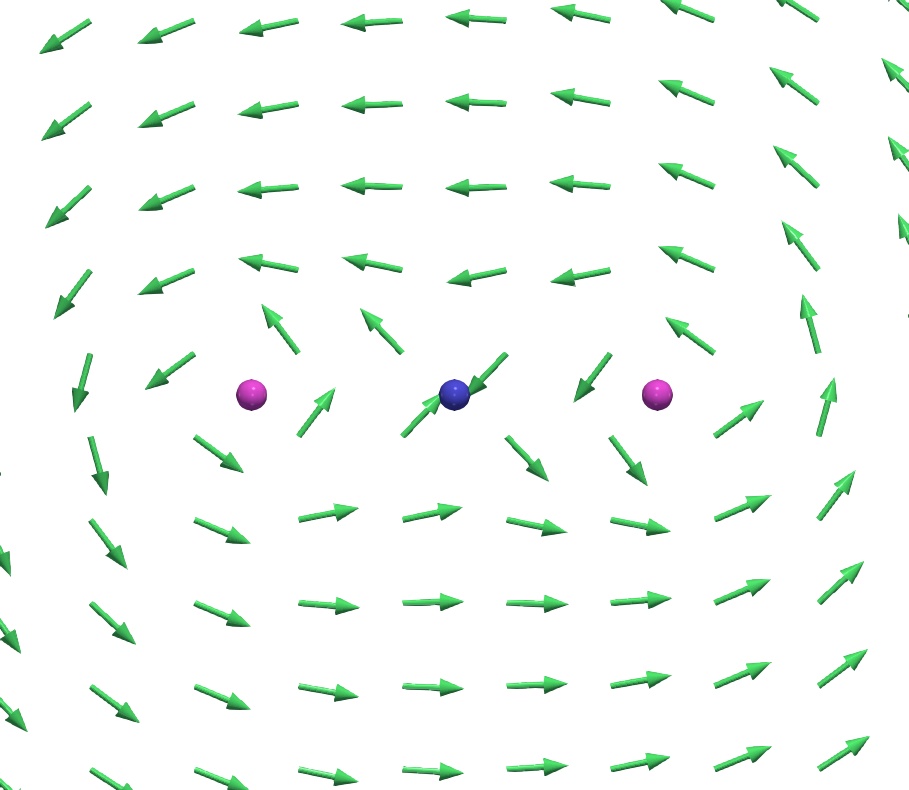}
    \end{minipage}
   	& \begin{minipage}{.3\linewidth}\vspace{0.1cm}
      \includegraphics[width=\linewidth]{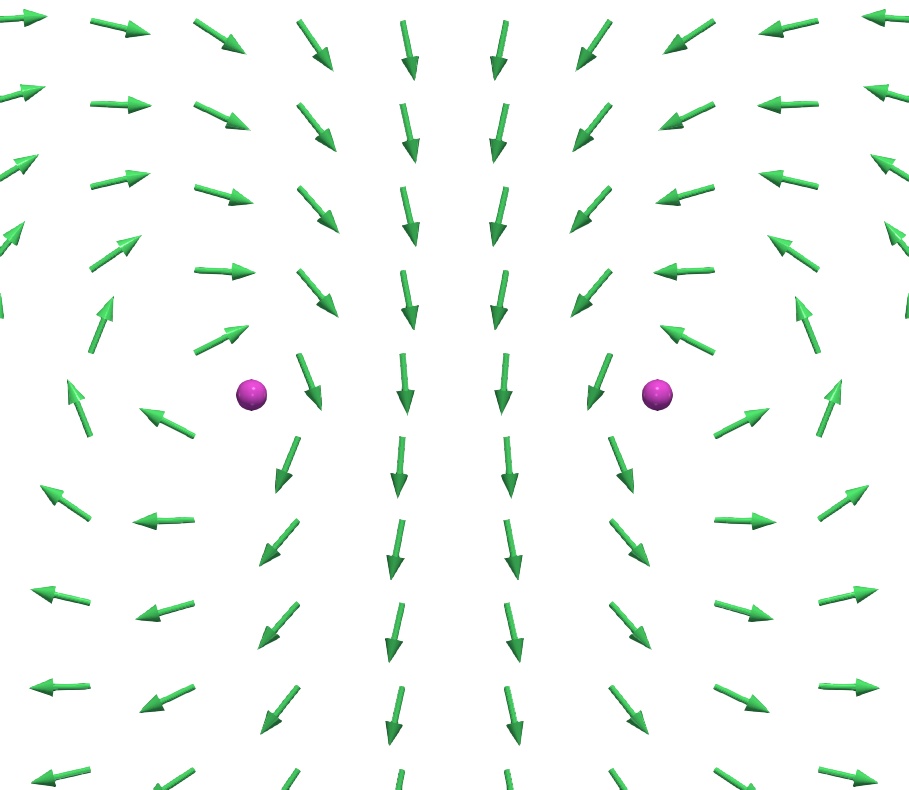}
    \end{minipage}  
   	\\ \hline
  
  \end{tabular}
  \caption{
	The Figures compare the difference between a vortex-vortex pair and a 
	vortex-antivortex pair, as fixed points of the XY-vector \Eqref{A-1} and 
	superflow vector field $\vv$ in \Eqref{v}. 
	The vortex and antivortex are respectively a center and a saddle, of the 
	XY-vector field $\vv_{\mathrm{XY}}$. On the other hand, they are both centers of 
	the 	vector field $\vv$.
	The red dots are the positions of the phase singularities. The blue dots show 
	the position of the velocity vector field saddle point. It is notable that the 
	saddle of the superflow is not associated with a phase singularity, but is a 
	stagnation point of $\vv$.
	}
  \label{Table:illustration:2}
\end{table}

In the Table~\ref{Table:illustration:2} we compare the vector fields $\vv_{\mathrm{XY}}$ 
\Eqref{A-1} and $\vv$ \Eqref{v} for a vortex-vortex pair, and for a vortex-antivortex 
pair. Unlike the XY-vector field \Eqref{A-1} for which there is no saddle-point in the 
case of a vortex-vortex pair, the superflow vector field $\vv$ has a saddle-point. 
Remarkably the phase portrait of $\vv_{\mathrm{XY}}$, in the case of a vortex-vortex 
pair, has the same topology as the vector field $\vv$ \Eqref{v} in the case of a 
vortex-antivortex pair. 
Moreover, the XY phase portrait in the case of a vortex-antivortex  pair
is similar to the vortex-vortex pair phase portrait of \Eqref{v}, but with 
one of the vortices (the one on the left in the Figure) moved to infinity.

\input{Poincare-index-BEC-final.bbl}

\end{document}

%% file: Poincare-index-BEC-final.bbl
\providecommand{\href}[2]{#2}\begingroup\raggedright\endgroup